\documentclass[%
 reprint,
 amsmath,amssymb,
 aps,
]{revtex4-2}

\usepackage{graphicx}
\usepackage{dcolumn}
\usepackage{bm}

\begin{document}

\preprint{APS/123-QED}

\title{Cavity optomechanical bistability with an ultrahigh \\reflectivity photonic crystal membrane}

\author{Feng Zhou}
 \altaffiliation[Also with ]{Joint Quantum Institute, NIST/University of Maryland, College Park, MD 20742, USA}
\author{Yiliang Bao}
\altaffiliation[Also with ]{Theiss Research, La Jolla, California 92037, USA}%
 \author{Jason J. Gorman}
 \author{John Lawall}
 \email{john.lawall@nist.gov}
\affiliation{%
 Microsystems and Nanotechnology Division, Physical Measurement Laboratory, National Institute of Standards and Technology, Gaithersburg, MD 20899, USA\\
}%

\date{\today}

\begin{abstract}
Photonic crystal (PhC) membranes patterned with sub-wavelength periods offer a unique combination of high reflectivity, low mass, and high mechanical quality factor. We demonstrate a PhC membrane that we use as one mirror of a Fabry-Perot cavity with finesse as high as $F=35\,000(500)$, corresponding to a record high PhC reflectivity of $R=0.999835(6)$. The fundamental mechanical frequency is 426 kHz, more than twice the optical linewidth, placing it firmly in the resolved-sideband regime. The mechanical quality factor in vacuum is~$Q=1.1(1)\times 10^6$, allowing us to achieve values of the single-photon cooperativity as high as ${\cal C}_0=6.6\times10^{-3}$. We easily see optomechanical bistability as hysteresis in the cavity transmission. As the input power is raised well beyond the bistability threshold, dynamical backaction induces strong mechanical oscillation above 1~MHz, even in the presence of air damping. This platform will facilitate advances in optomechanics, precision sensing, and applications of optomechanically-induced bistability.
\end{abstract}

\maketitle

\section{Introduction}
It has long been recognized that patterning a dielectric membrane with a 1D array of lines or a 2D array of holes can dramatically increase its optical reflectivity.  In the 1D case, such devices are often called ``high contrast gratings,’’ \cite{chang-hasnain_high-contrast_2011} and their remarkable optical and mechanical properties have been exploited to realize high-finesse optical cavities \cite{bruckner_realization_2010,kemiktarak_mechanically_2012}, high-Q monolithic optical resonators\cite{zhou_surface-normal_2008}, second harmonic emission \cite{tran_surface-normal_2013}, tunable vertically-coupled surface-emitting lasers (VCSELs) \cite{huang_surface-emitting_2007}, optical cooling \cite{kemiktarak2012cavity}, and phonon lasing \cite{kemiktarak_mode_2014}.  In the 2D case, such ``photonic crystal slabs’’ \cite{PhotonicXtals} or ``PhC membranes’’ have found numerous applications in both optics \cite{lousse_angular_2004,singh_chadha_polarization-_2013,deng_modeling_2013,krishnan_enhanced_2016,krishnan_tunable_2018} and optomechanics \cite{antoni_deformable_2011,norte_mechanical_2016,chen_high-finesse_2017,gartner_integrated_2018,morsy_high_2019,lien_experimental_2022,xu_millimeter-scale_2022}.  For cavity optomechanical applications, a key property of these devices is the reflectivity that is achieved, which determines the finesse of the optical cavity in which it is employed.  Early work with 1D structures  demonstrated cavities with finesse $F = 1\,200$ to $F = 2\,800$ \cite{kemiktarak2012cavity,kemiktarak_mechanically_2012},  and subsequent ``membrane-in-the-middle’’ \cite{stambaugh_membrane---middle_2015} experiments showed that even higher values of the finesse were possible.  More recently, a finesse of $F=6\,390(150)$ was demonstrated in a SiN structure patterned with a 2D square lattice of holes \cite{chen_high-finesse_2017}.  In this work, we demonstrate a record high finesse of $F=35\,000(500)$, corresponding to a PhC reflectivity of $R=0.999835(6)$, using a PhC membrane with a hexagonal lattice. 
We find that the wavelength dependence of the finesse is Lorentzian, 
and elucidate the functional form of the global transmission spectrum.

From the beginning of this study, 
the phenomenon of optomechanical bistability, arising from radiation pressure elongating the static cavity length, has played a key role.  Bistability resulting from this Kerr-type 
nonlinearity was first observed \cite{dorsel_1983} via hysteresis in the power transmitted by a mechanically compliant cavity as the drive power was swept. This pioneering experiment was very sensitive to seismic noise, and
the rare successful experiments took place only at night.  
Following the initial demonstration, a great deal of interest in optomechanical bistability was driven by the recognition that the underlying Kerr-like nonlinearity does not suffer from the material losses limiting traditional Kerr media \cite{meystre_theory_1985,miri_optomechanically_2017}.  These ideas have motivated proposals for nonreciprocal devices \cite{xu_optomechanically_2018}, all-optical switches and  memories based on spontaneous symmetry breaking \cite{miri_optomechanically_2017}, and devices with negative effective mass \cite{prakash_negative_2019}.
For sensing, it has been shown thoeretically that bistable optomechanical systems provide force sensitivity superior to that of their linear counterparts \cite{aldana_detection_2014}, 
and in the quantum domain, ponderomotive squeezing \cite{FabreSqueezing1994} and entanglement \cite{ghobadi_quantum_2011} are enhanced near the bistable critical points. Very recent work in a magnetic system described by the same Kerr nonlinearity has shown enhanced optomechanical cooling by exploiting the asymmetric lineshape characteristic of the system below the bistability threshold \cite{zoepfl_kerr_2022}.  Even the degree to which the optomechanical system is equivalent to a Kerr medium has been the subject of a detailed analysis \cite{aldana_equivalence_2013}.

Despite the numerous proposals for experiments and applications, very few other optomechanical systems exhibiting radiation-pressure induced bistability or multistability have been reported \cite{mueller_observation_2008,rochau_dynamical_2021,xu_observation_2017}.  
In some cases, reports of bistability have involved photothermal mechanisms including expansion of the mirror coatings \cite{altin_robust_2017,noauthor_photothermally_2020} and intensities as high as ${\rm 3\,MW/cm^2}$ \cite{ma_observation_2020}.  Here, the low mass and high reflectivity of our PhC membrane enable an optomechanical cavity with a fundamental mechanical frequency of 426~kHz in which hysteresis induced by bistability is easily seen with incident optical powers below $500\,\mu$W.  We present a quantitative discussion of our observations, which include the appearance of oscillation (instability of the upper branch \cite{aldana_equivalence_2013}) as the power is raised, despite the damping provided by air. 
While we focus on static bistability in the present study, the mechanical quality factor of our device in vacuum is $Q=1.1(1)\times 10^6$, which, in combination with its low mass and ultrahigh reflectivity, allows us to achieve a single-photon cooperativity \cite{aspelmeyer_cavity_2014} as high as ${\cal C}_0=6.6\times10^{-3}$.  This renders our device attractive for dynamic applications as well, especially where the simplicity of a two-mirror cavity is desired.

\section{Experimental setup}
\subsection{Photonic crystal membrane}

\begin{figure}[t!]
\centering\includegraphics[width=1\linewidth]{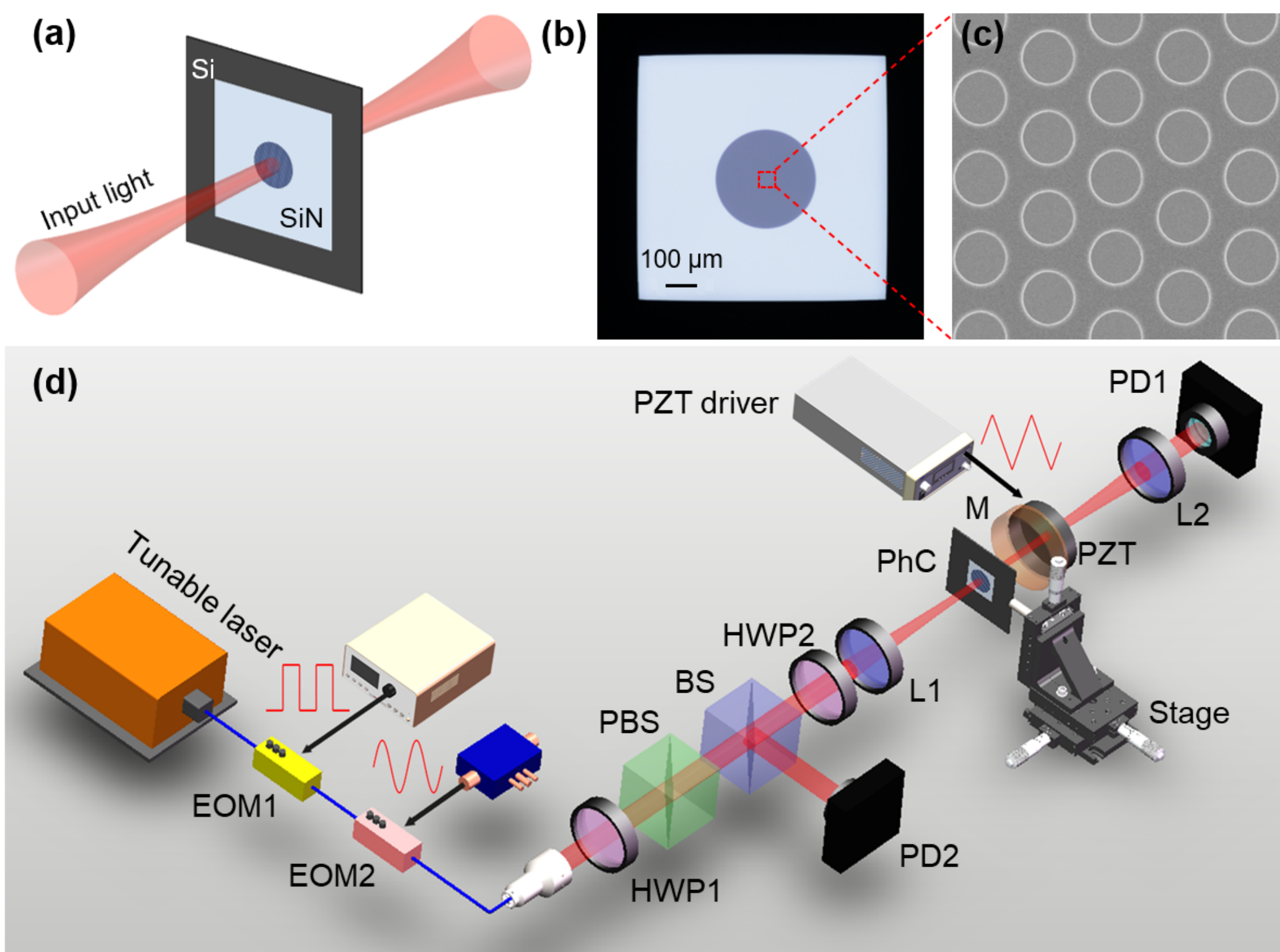}
	\caption{\label{setupfig1} Photonic crystal design and experimental setup for characterization. {\bf (a)} A silicon nitride (SiN) membrane is suspended by a Si frame, which has a patterned PhC disk at the center. A focused laser beam is  normally incident. {\bf (b)} An optical image of the photonic crystal chip with an $800\,\mu$m square SiN membrane (light blue), a $300\,\mu$m diameter photonic crystal, and a Si frame (black). {\bf (c)} Scanning electron micrograph of the photonic crystal, showing a hexagonal lattice of circular holes. {\bf (d)} Experimental setup used to characterize properties of the photonic crystal. EOM1: electro-optic intensity modulator; EOM2: electro-optic phase modulator; HWP1, HWP2: half-wave plates; PBS: polarizing beam splitter; BS: beam splitter; L1, L2: lenses (f = 200 mm); PhC: photonic crystal membrane; M: high-reflectivity concave mirror; PD1, PD2: photodetectors.} 
\end{figure}

Our photonic crystal membrane is illustrated in Fig.~\ref{setupfig1}(a).  It is a square silicon nitride (SiN) membrane, 220~nm in thickness and 800~$\mu$m on a side, suspended in a silicon  frame.  A 2D PhC of diameter 300~$\mu$m is patterned in the center.  Figure~\ref{setupfig1}(b) shows an optical image of the PhC membrane, and Fig.~\ref{setupfig1}(c) is a scanning electron micrograph of the center.  The PhC is made of a hexagonal lattice of circular holes etched into the SiN device layer.  Its geometry was chosen to maximize the reflection of a normally incident Gaussian beam at 1550~nm using $S^4$~\cite{Liu20122233}, a freely available software package employing the Rigorous Coupled Wave Analysis (RCWA) algorithm.

Fabrication started by coating a $525\,\mu$m thick silicon wafer with a $220$~nm layer of low-stress SiN using low pressure chemical vapor deposition (LPCVD). The photonic crystals were patterned with electron-beam lithography and the silicon nitride was etched with reactive ion etching (RIE). The backside of the wafer was then patterned with square openings to form silicon nitride membranes by fully etching through the silicon wafer. Deep reactive ion etching (DRIE) was used to etch the back side of the silicon wafer until approximately $50\,\mu$m of silicon was left. After dicing into 1~cm square chips, the membranes were fully released using KOH with a concentration of 30\,\% at $60~^{\circ}$C.  In this fabrication process, the lattice constant can be controlled precisely, whereas the hole radius is much more prone to process variations. Therefore, we fabricated multiple PhCs with the same lattice constant ($a = 1.51\,\mu$m) but different target radii (varying from $r = 0.475\,\mu$m to $r = 0.515\,\mu$m; simulations gave an optimum radius of $r = 0.515\,\mu$m).  An alternative approach involving iterative etching to approach a target wavelength has been demonstrated elsewhere~\cite{bernard_precision_2016}.

\subsection{Optical Interrogation}
The experimental apparatus is shown in Fig.~\ref{setupfig1}(d).  A Fabry-Perot cavity is created by using the PhC membrane as the input coupler, and a concave dielectric mirror as the output coupler.  The output coupler has a nominal radius of curvature of 25~mm and a transmission (as documented by the manufacturer) of $T_2 = 10^{-5}$. The PhC membrane is mounted on a translation stage allowing alignments in 3D, and the output coupler is mounted by means of a piezoelectric transducer on a 1D translation stage enabling motion along the axis of the cavity.  
Laser light from a narrow-linewidth tunable laser 
is collimated and mode-matched with a lens (f = 200 mm) into the cavity. A half-wave plate and a polarizing beamsplitter are used to adjust the input power. The transmission of a hexagonal photonic crystal lattice is not polarization-independent, so the polarization of the incident light is optimized by means of another half-wave plate.
Transmitted light is collimated by a lens (f = 200mm) and captured by a photodetector, while the reflection is detected by another photodetector. An electro-optic intensity modulator allows the incident light to be rapidly extinguished for cavity ringdown measurements, and an electro-optic phase modulator is used to introduce sidebands to the laser frequency spectrum. These sidebands can either be driven by a synthesizer to establish a precise frequency calibration, or driven by a voltage-controlled oscillator in order to rapidly sweep the optical frequency.

\begin{figure*}[ht]
	\centering
	\includegraphics[width=0.92\linewidth]{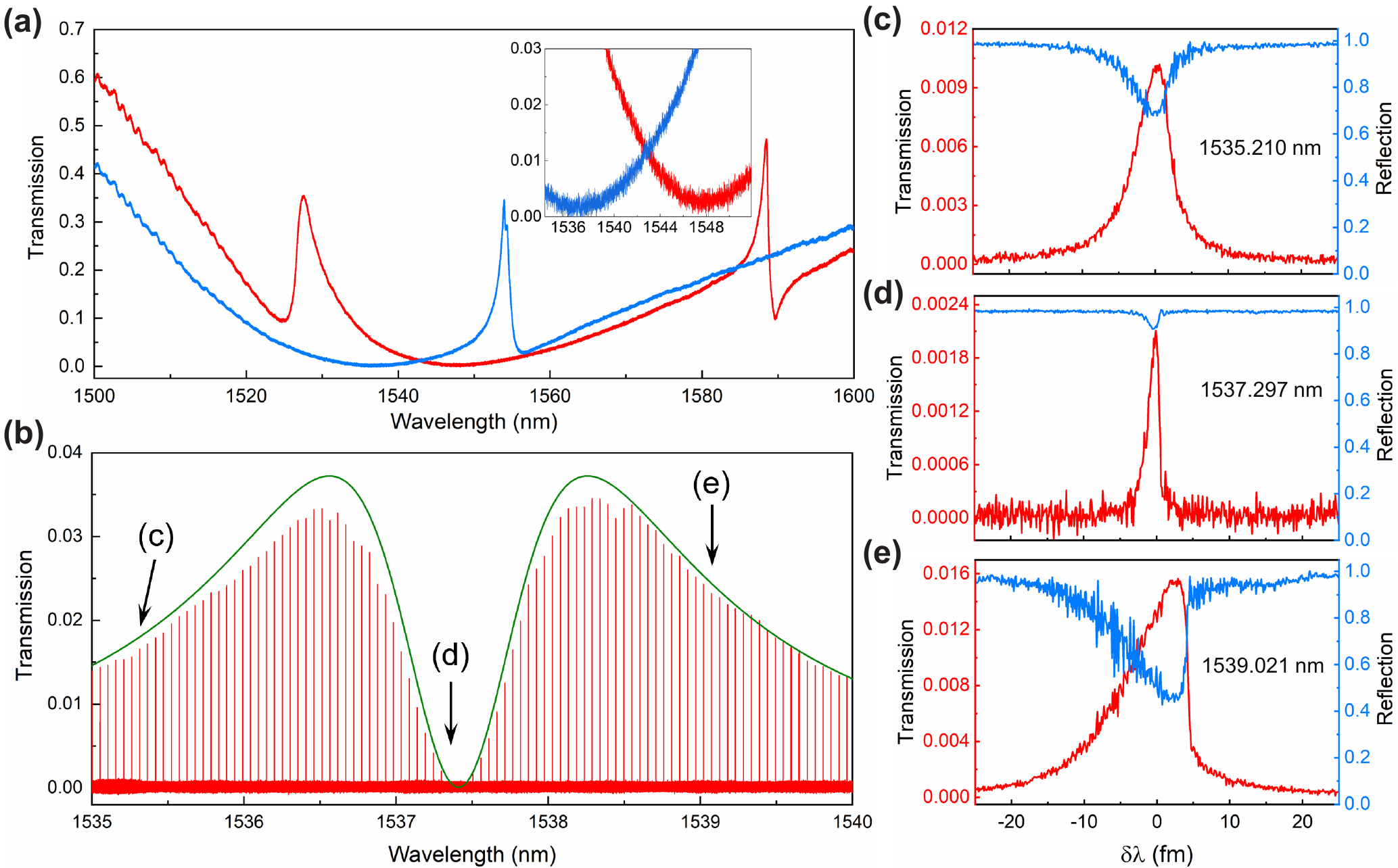}
	\caption{\label{CavitySpectra1} Optical characterization of the photonic crystal membrane. {\bf (a)} Transmission spectra of two different devices for light focused to a spot size of 60 µm. Transmission minima are exhibited at around 1537 nm (blue) and 1548 nm (red), corresponding to devices with design hole radii of 505 nm and 475 nm respectively. The inset shows a zoomed-in view, indicating a slightly lower transmission minimum at 1537 nm. We focus on this device in the following measurements. {\bf (b)} Transmission spectrum for a ``long'' cavity made using an input optical power of 700 µW. The green line is a theoretical transmission envelope calculated from the measured wavelength dependence of the cavity finesse. {\bf (c)}, {\bf (d)}, {\bf (e)} Zoomed-in views of modes at 1535.210~nm, 1537.297~nm and 1539.021~nm, respectively, with both transmission (red) and reflection (blue) signals displayed. The resonance at 1537.297~nm has considerably narrower linewidth. The lineshapes of the modes are asymmetric, with steeper slopes on the red-detuned sides, due to optomechanical nonlinearity.}
\end{figure*}

\section{Optical characterization}
\subsection{Photonic crystal membrane transmission spectrum}
Initial measurements of the photonic crystal properties were made in the absence of the cavity output coupler by simply focusing light onto the sample and monitoring the transmission as a function of wavelength.  Transmission spectra are shown in Fig.~\ref{CavitySpectra1}(a) for two of the eight devices on a sample. Both spectra exhibit minima in transmission along with resonant features that are well known to occur in such devices~\cite{bernard_precision_2016}. While the spectrum of the device with design radius $r=475$~nm (red) has its minimum very close to the target wavelength of 1550~nm, the device with $r=505$~nm (blue) has a slightly lower minimum transmission, as shown in the inset, and is the one studied in the remainder of this work.  

\subsection{Cavity transmission spectrum}
We then installed the concave dielectric output coupler forming a Fabry-Perot cavity, and measured the transmission and reflection of the cavity near the wavelength that minimizes the transmission of the photonic crystal. We optimized the cavity finesse by working with a cavity length such that the cavity waist is $\omega_0\approx 60\,\mu$m, obtained by taking either 
$L \approx 2.4$~mm (``short'' cavity) or $L \approx 22.8$~mm (``long'' cavity).  The optimal cavity waist reflects a compromise between the limited size of the PhC and the broader angular spectrum of a more tightly focused optical mode, as addressed in previous work~\cite{kleckner2010diffraction,kemiktarak2012cavity}. As we will show, both choices of cavity length give the same finesse, but the free spectral range and linewidth differ by about a factor of ten in the two configurations.
A transmission spectrum made with a ``long'' cavity, driven with incident power of $P_{in}=700\,\mu$W, is shown in Fig.~\ref{CavitySpectra1}(b).  It exhibits  approximately 100 modes over the range of 1535~nm to 1540~nm. The cavity transmission varies dramatically over this range, exhibiting maxima at 1536.5~nm and 1538.2~nm, and dropping to nearly zero in between; we will discuss the transmission envelope in Sec.~4B. Profiles of both transmission and reflection for three of the modes are shown in Fig.~\ref{CavitySpectra1}(c)-(e). It is clear that the resonance at 1537.297~nm (Fig.~\ref{CavitySpectra1}(d)) is considerably narrower than the others; in addition, all three modes are asymmetric, with a steeper slope on the long-wavelength (red-detuned) side of the resonance. As we will discuss in Sec.~\ref{sec: optomechanics}, the lineshape asymmetry is due to optomechanical nonlinearity. 

\begin{figure*}[t!]
	\centering
	\includegraphics[width=0.85\linewidth]{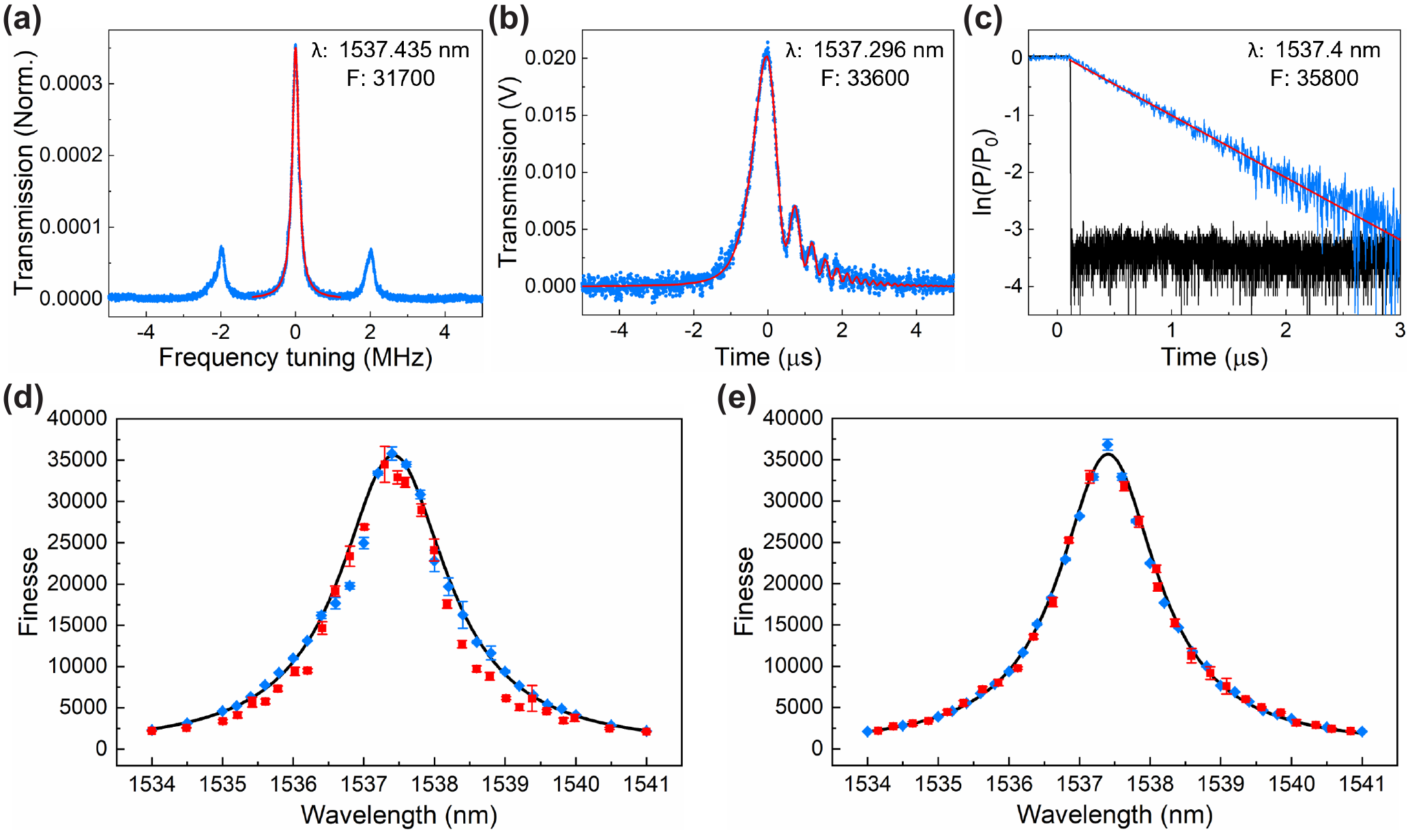}
	\caption{\label{Finesse1}Cavity finesse measurements. {\bf (a)} An Airy transmission resonance normalized to the input power at 1537.435 nm with 2 MHz sidebands for calibration, along with a Lorentzian fit (red).  The measurement is made at very low power to minimize optomechanical nonlinearities. {\bf (b)} Measured transmission with a phase modulator sideband swept rapidly across a resonance peak centered at 1537.296 nm.  A theoretical profile (red) is fit to the data. {\bf (c)} Cavity ringdown measurement. The decay signal is displayed on a natural logarithmic scale (blue) with a linear fit (red). The black curve shows extinction of the excitation light.  {\bf (d)} The measured finesse as a function of wavelength using the cavity ringdown method (blue diamonds) and fast sweeping method (red squares) for a ``long''  cavity. The ringdown data are fit to a Lorentzian lineshape (black line). Error bars are given by the standard deviation of five individual measurements.  {\bf (e)} Same as {\bf (d)}, but for a ``short'' cavity.}  
\end{figure*}

In general the asymmetric lineshapes preclude fitting to a Lorentzian functional form.  A brute-force approach to determining the finesse from the linewidth is to reduce the optical power to the point that the optomechanical distortion in the resonances is negligible.  This approach is demonstrated in Fig.~\ref{Finesse1}(a), which shows the transmission at a wavelength of $\lambda = 1537.435$~nm of a ``long’’ cavity with a free spectral range of FSR = 6.83(5)~GHz. An electro-optic phase modulator  is used to generate sidebands at a separation of 2~MHz for frequency calibration. The linewidth (FWHM) of the central resonance is found to be 215(16)~kHz, corresponding to a finesse $F=31\,900(2500)$, where the uncertainties correspond to the standard deviation of a series of 14 measurements.
This approach clearly introduces the laser linewidth as a broadening mechanism.  In addition, it requires one to sweep the laser frequency relatively slowly in order to obtain an adequate signal/noise ratio, so that the measurement is also susceptible to cavity length fluctuations. We thus employed alternative means to determine the cavity finesse.

\subsection{Frequency-swept measurements}
A more desirable method to determine the cavity finesse in this case
is to sweep the laser wavelength rapidly through the resonance.
The transmission of a cavity whose length is rapidly changed~\cite{poirson_analytical_1997}, or that is driven by an input field whose frequency is rapidly swept, exhibits
secondary oscillations (ringing), whose functional form can be used to infer the cavity finesse.  As discussed previously \cite{poirson_analytical_1997}, this approach is particularly well suited to high-finesse cavities.
In the present work, the method presents two advantages. First, the light is swept through the resonance so quickly that very little circulating power builds up, reducing the associated nonlinearity induced by radiation pressure. Second, the approach mitigates the effect of technical noise related to environmental disturbances of the cavity length, once again because the measurement is done so quickly.

We perform the measurements by sweeping the drive frequency of the electro-optic phase modulator at a rate~$\beta$ ($8\times 10^6$~MHz/s < $\beta$ < $ 10^8$~MHz/s) such that one sideband scans through a cavity resonance, and monitoring the cavity transmission.  
The transmitted power $P_{trans}(t)$ is proportional to (see Supplementary Material)
\begin{equation}
P_{trans}(t)\propto e^{-2\gamma t}
\left|{\rm erfc}\left(\frac{1}{2}\frac{(1+i)(\gamma-i\beta t)}{\sqrt{-\beta}}\right)\right|^2,
\label{eqn: usquared}
\end{equation}
where $\gamma$ is the cavity half-width at half maximum (HWHM) in units of angular frequency. 
Figure~\ref{Finesse1}(b) shows a measured transmitted signal (blue) for $\lambda=1537.296$~nm and $\beta=8.8(3)\times 10^6$~MHz/s, along with a fit (red) to~Eq. (\ref{eqn: usquared}). The linewidth inferred is $2\gamma=2\pi\times 208(20) $~kHz, corresponding to a finesse $F=32\,800(3000)$, where the uncertainties correspond to the standard deviation of five successive measurements.

\subsection{Ringdown measurements}
A complementary approach to measuring optical cavity losses is to work in the time domain and study cavity ringdown.  Here light is coupled into the cavity and then abruptly extinguished, and the light stored in the cavity leaks out with a time constant given by $\tau=1/(2\gamma)$.  
This approach is immune to broadening due to the laser linewidth, as well as optomechanical effects and environmental perturbations in the cavity length over the course of the measurement.
Incoupled light was intensity modulated with a square wave, and the power exiting the output coupler was measured with a fast photodetector (3~ns risetime).
Ringdown data for a cavity with a free spectral range of 6.58~GHz at a wavelength of $\lambda=1537.4$~nm is shown in Fig.~\ref{Finesse1}(c) on a logarithmic scale. From a linear fit we infer $\tau=874(36)$~ns, corresponding to a linewidth of $2\gamma=2\pi\times183(8)$~kHz and a finesse of $F=35\,900(1500)$; once again the uncertainty is given by the standard deviation of five consecutive measurements. 

\section{Implications: Wavelength dependence of finesse and transmission}
\subsection{Wavelength dependence of finesse}
Figure~\ref{Finesse1}(d) and Fig.~\ref{Finesse1}(e) show the cavity finesse inferred by the fast sweep and ringdown approaches for both the ``long'' and ``short'' cavity configurations, respectively.  The range of wavelengths covered in these graphs is very close to the minimum of the transmission curve shown in Fig.~\ref{CavitySpectra1}(a), for which one can expand the transmission of the photonic crystal in a Taylor series as a function of wavelength as
\begin{equation}
    T(\lambda)=T_1^0+\alpha(\lambda-\lambda_0)^2+...
    \label{eqn: Tvslambda}
\end{equation}
The finesse of a low-loss cavity is given by
\begin{equation}
    F=\frac{2\pi}{\Sigma \,roundtrip\,\,losses}
\end{equation}
where the ``roundtrip losses'' in the denominator refer to mirror transmission as well as undesired (scattering and absorption) losses.  Denoting the total absorption and scattering losses by $A$ and $S$, respectively, the wavelength dependence of the finesse in our case is, in the vicinity of~$\lambda_0$,
\begin{equation}
    F(\lambda)=\frac{2\pi}{A+S+T_2+T_1^0+\alpha(\lambda-\lambda_0)^2}
    \label{eqn: finesse}.
\end{equation}
Here it is implicit (as documented in the spectrophotometer data provided by the manufacturer) that the transmission of the output coupler $T_2$ is independent of wavelength for the wavelengths of interest here. Eq.~(\ref{eqn: finesse}) has the functional form of a Lorentzian; fitting Lorentzians to the ringdown data shown in Fig.~\ref{Finesse1}(d) and~\ref{Finesse1}(e) yields a peak finesse
$F_{peak}=35\,000(500)$, and a corresponding determination of the total cavity losses as
\begin{equation}
    A+S+T_2+T_1^0=1.80(3)\times 10^{-4}.
    \label{eqn: losses}
\end{equation}
Taking the known $T_2 = 10^{-5}$, we make the reasonable assumption that the absorption and scattering losses of the dielectric mirror are comparable to, but somewhat smaller than, its transmission, so that $T_2+A_2+S_2=1.5(5)\times 10^{-5}$.  We can then infer for the photonic crystal
\begin{equation}
    A_{PhC}+S_{PhC}+T_1^0=1.65(6)\times 10^{-4}.
    \label{eqn: losses2}
\end{equation}
or, equivalently, a peak reflectivity $R_{PhC}=0.999835(6)$.  

\subsection{Wavelength dependence of transmission}
\label{sec: envelope}
We now address the envelope of the transmission spectrum shown in Fig.~\ref{CavitySpectra1}(b).  The resonant transmission of a Fabry-Perot cavity is given by
\begin{equation}
    T_{cav}=\eta\left(\frac{F}{\pi}\right)^2T_1T_2
    \label{eqn: Tcav0},
\end{equation}
where $\eta$ is a numerical factor $(0\le\eta\le 1)$ representing the degree of mode-matching. Combining Eq.~(\ref{eqn: Tvslambda}),~(\ref{eqn: finesse}), and~(\ref{eqn: Tcav0}), we obtain
\begin{equation}
    T_{cav}=4\eta T_2\frac{T_1^0+\alpha(\lambda-\lambda_0)^2}{\left(A+S+T_2+T_1^0+\alpha(\lambda-\lambda_0)^2\right)^2}
    \label{eqn: Tcav}.
\end{equation}
Noting that $A+S+T_2+T_1^0$ and $\alpha$ have already been determined from the fits in Fig.~\ref{Finesse1}(d,e) , the only free parameters are $T_1^0$ and $\eta$.  The value of $T_1^0$ plays little role here; tentatively taking $T_1^0=0$ and $\eta=0.64$, we find the envelope shown in green in Fig.~\ref{CavitySpectra1}(b).  The qualitative agreement is quite good, and shows that the shape of the global transmission spectrum is set by competition between the finesse (favoring a small $T_1(\lambda))$ and the input coupling (favoring a large $T_1(\lambda)$).  

A more refined value of $T_1^0$ can be found by using transmission data such as that shown in Fig.~\ref{Finesse1}a in conjunction with Eq.~(\ref{eqn: Tcav0}).  Indeed, at $\lambda=\lambda_0$, we can infer from Eq.~(\ref{eqn: Tcav0})
\begin{equation}
    T_1^0=\frac{T_{cav}(\lambda_0)}{\eta T_2}\left(\frac{\pi}{F_{max}}\right)^2
    \label{eqn: T10}.
\end{equation}
From a series of 14 measurements like the one in Fig.~\ref{Finesse1}a, we infer a cavity transmission $T_{cav}(\lambda_0)=3.7(4)\times10^{-4}$, where the uncertainty is given by the standard deviation of the measurements, and a corresponding value $T_1^0=5(1)\times 10^{-7}$.  It is clear from comparison to~(\ref{eqn: losses2}) that absorption and scattering losses dominate the transmission at the wavelength~$\lambda_0$.  In fact, the transmission $T_2$ of the dielectric output coupler is an order of magnitude larger than that of the photonic crystal at this wavelength, so when operating at maximum finesse, it may be advantageous to couple light in through the curved dielectric mirror rather than the photonic crystal.
\subsection{Optimal power buildup and impedance matching}
The dependence of the finesse and transmission on wavelength, governed by the reflection spectrum of the photonic crystal, allows the properties of the cavity resonances to be chosen for the task at hand.  In some cases a very narrow resonance might be desired, and it would be advantageous to choose a wavelength close to the one maximizing the finesse.  In other cases, however, it might be preferable to work at a wavelength that maximizes the circulating optical power.  Differentiating Eq.~(\ref{eqn: Tcav}), one finds that the 
value of $T_1(\lambda)$ maximizing the transmission is
\begin{equation}
    T_1(\lambda)=A+S+T_2.
\end{equation}
This condition, in which the transmission of the input coupler equals the sum of all other round-trip losses, is well known to define an impedance-matched cavity \cite{Siegman}, which, in the case of perfect mode-matching, has a vanishing resonant reflection. Indeed, the reflection dip shown in Fig.~\ref{CavitySpectra1}(e), while not at the exact wavelength of impedance matching, is substantially deeper than those of Fig.~\ref{CavitySpectra1}(c) and Fig.~\ref{CavitySpectra1}(d).  In the limit $T_1^0\rightarrow 0$, the wavelength detuning yielding this impedance-matched situation is equal to the half-width at half maximum of the Lorentzian function (Fig.~\ref{Finesse1}(d) and~\ref{Finesse1}(e)) describing the variation of the finesse with wavelength.  

\section{Optomechanical bistability and dynamics at atmospheric pressure}
\label{sec: optomechanics}
We return now to the asymmetric lineshapes shown in Fig.~\ref{CavitySpectra1} (c), (d), (e). When optical power is stored in a Fabry-Perot cavity, the associated radiation pressure elongates the cavity and shifts its resonance frequencies down.  While this effective Kerr nonlinearity is in theory always present, it is only apparent if the cavity is sufficiently mechanically compliant.  By balancing the mechanical restoring force against the radiation pressure, the steady-state circulating power is found to obey a cubic equation:
\begin{equation}
P_{circ}=\frac{P_{max}}{\left(\frac{\delta}{\gamma}+\frac{P_{circ}}{\tilde{P}}\right)^2+1} 
\label{eqn: Pcirc}
\end{equation}
where
\begin{equation}
\tilde{P}=\frac{k_{static}c\lambda}{8F}
\label{eqn: Ptilde}
\end{equation}
is the circulating power needed to shift the resonant frequency of the cavity to the red by one half-linewidth~$\gamma$, and $P_{max}$ is the maximum possible circulating power, dependent on the detuning~$\delta$ and mode-matched input power. Here~$\delta$ is defined relative to an undriven cavity, where there is no radiation pressure, and $k_{static}$ is the mechanical spring constant of the optical cavity.  In addition to the static nonlinearity discussed here, it is well known that the optomechanical interaction modifies the system dynamics \cite{aspelmeyer_cavity_2014}, and the individual mechanical modes have their own effective masses and corresponding spring constants.

\begin{figure}[ht]
	\centering
	\includegraphics[width=0.95\linewidth]{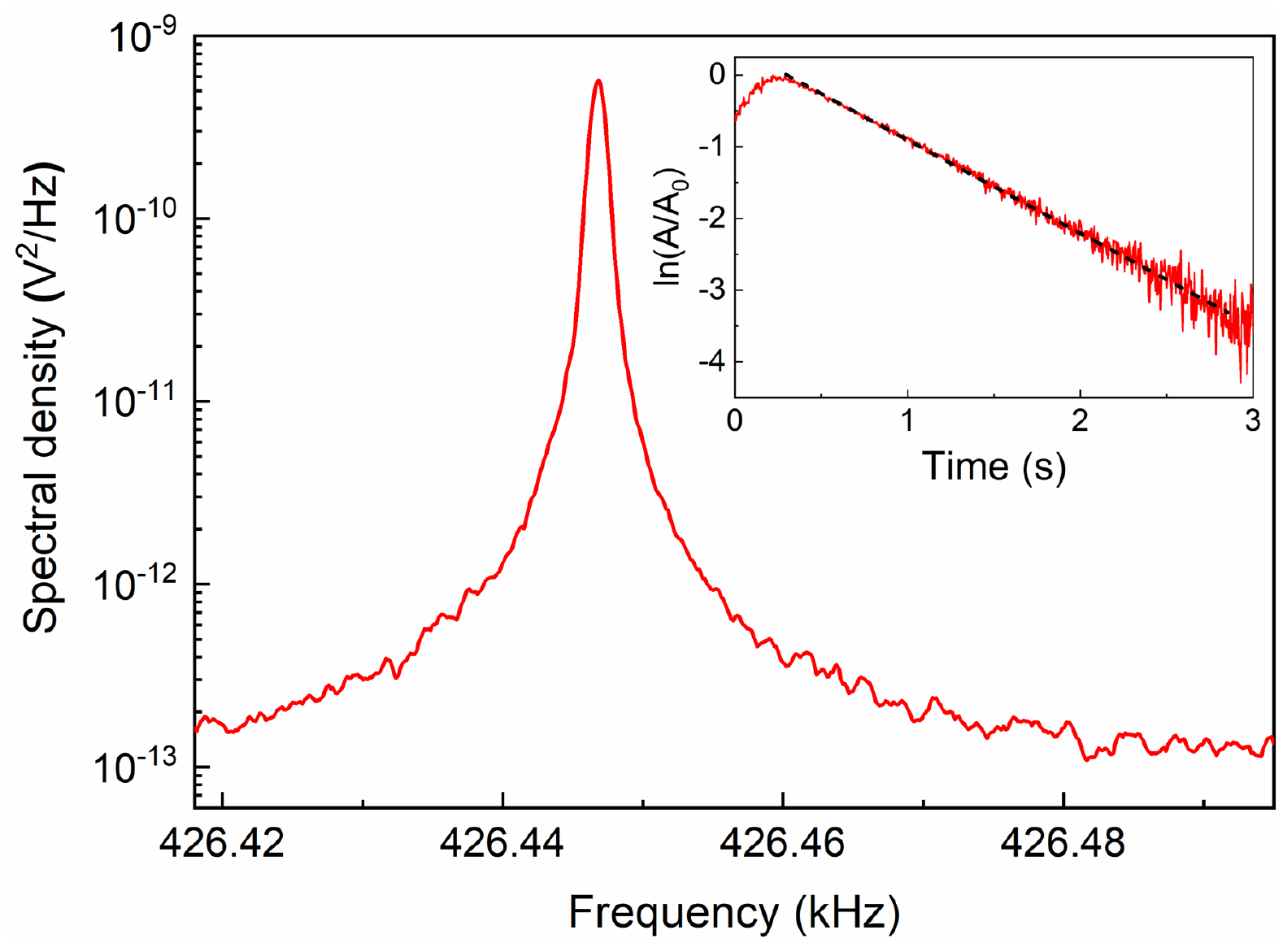}
	\caption{\label{thermal}Thermomechanical spectrum of the PhC membrane measured in vacuum with a Michelson interferometer. The inset shows a ringdown measurement of the oscillation amplitude~$A$, from which we infer a quality factor ~$Q=1.1(1)\times 10^6$. }
\end{figure}

The fundamental frequency of our photonic crystal membrane is $\nu_m\approx 426$~kHz, as illustrated by the power spectral density shown in Fig.~\ref{thermal}. This measurement of the thermal Brownian motion was made using a Michelson interferometer in a vacuum environment.  The inset shows a ringdown measurement from which we infer a mechanical quality factor of~$Q=1.1(1)\times 10^6$, where the uncertainty corresponds to the standard deviation of three measurements.  For radiation pressure from a Gaussian beam with spot size $\omega_0=60\,\mu$m, the static spring constant of our membrane can be expressed in terms of the fundamental frequency $\omega_m=2\pi\nu_m$ of the membrane as $k_{static}\approx 0.089m\omega_m^2$ (see Supplementary Material), enabling one to calculate $\tilde{P}$ for any value of the finesse.

\begin{figure*}[ht]
	\centering
	\includegraphics[width=0.85\linewidth]{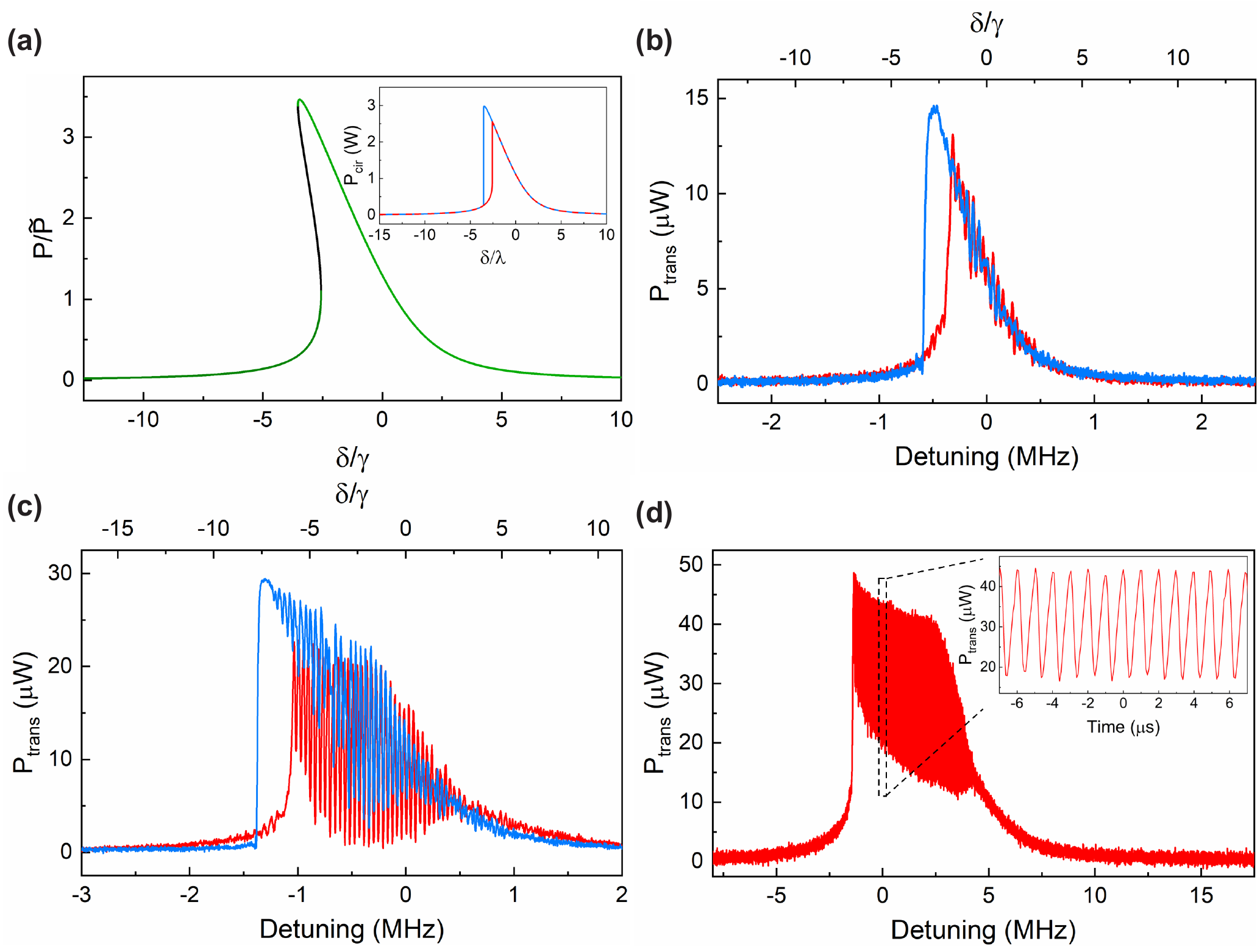}
	\caption{\label{hysteresis}{Hysteresis related to optomechanical bistability.  All detunings are relative to an undriven cavity; the detuning relative to the optomechanically elongated cavity is blue-shifted by $\frac{P_{circ}}{\tilde{P}}\gamma$.  \bf (a)} ``Shark fin'' solution to~(\ref{eqn: Pcirc}) for the circulating power as a function of detuning, for $P_{max} = 2\sqrt{3}\tilde{P}$. The solution  shown in black is physically unstable. Inset: Associated hysteretic behavior for sweeps from red to blue detuning (red curve), and from blue to red detuning (blue curve). The circulating power is expressed in Watts by means of $\tilde{P}\approx860$~mW appropriate to a finesse $F=18\,500$. {\bf (b)} Cavity transmission for a bidirectional laser sweep with an input power of $P_{in}=500\,\mu$W at a wavelength of $\lambda=1536.576$~nm. The color convention is the same as that taken for the inset to \ref{hysteresis}{\bf (a)}.  {\bf (c)} Bidirectional sweep when the laser input power is raised to 1~mW; the hysteretic behavior is still clearly observed, but it is accompanied by significant oscillation on the blue-detuned side of the resonance. {\bf (d)} At still higher input powers, the dynamical behavior dominates.  Here the cavity is driven with 4~mW at a wavelength of $\lambda= 1539.594$~nm and the detuning is swept from red to blue; the inset shows that the optically-sprung oscillation reaches a frequency slightly in excess of 1~MHz.} 
\end{figure*}

For $\tilde{P}\rightarrow\infty$, corresponding to a perfectly rigid cavity, Eq.~(\ref{eqn: Pcirc}) gives the usual Lorentzian response associated with a driven Fabry-Perot cavity.  With a finite spring constant, however, the Lorentzian becomes distorted, and when $P_{max}$ exceeds a threshold power $P_{th} = 8\sqrt{3}/9 \tilde{P}$, the system becomes bistable and hysteretic.  Figure~\ref{hysteresis}(a) shows the ``shark fin'' solution to Eq.~(\ref{eqn: Pcirc}) for $P_{max} = 2.25 P_{th}=2\sqrt{3}\tilde{P}$, in which the maximum circulating power for a detuning sweep from blue to red is 17\,\% larger than the maximum when the detuning is swept from red to blue. For values of the detuning $-3.54\,\gamma<\delta<-2.57\,\gamma$, Eq.~(\ref{eqn: Pcirc}) admits three real solutions, but only the two depicted in green are physically stable.  Sweeping the detuning from red to blue or from blue to red gives different responses, as illustrated in the inset.

Figure~\ref{hysteresis}(b) shows bidirectional spectra of the power transmitted by the cavity when driven with an input power of $P_{in}=500\,\mu$W at a wavelength of $\lambda=1536.576$~nm, where the finesse is $F\approx18\,500$.  The spectra are in good qualitative agreement with those shown in Fig.~\ref{hysteresis}(a) ; in particular, the transmitted power is about 17\,\% larger when sweeping from blue to red detuning than when sweeping from red to blue.  It is also notable that there is a small amount of oscillation on the blue-detuned side of the resonance, regardless of the direction of the detuning sweep.  This is the usual consequence of dynamical backaction \cite{aspelmeyer_cavity_2014}.  It is related to the fact that the upper branch of the optomechanical system can be unstable, unlike the case for a true Kerr medium, as discussed theoretically \cite{aldana_equivalence_2013}. 

More quantitatively, the physical mass of our membrane is $m=4.2\times10^{-10}$~kg, yielding a 
power 
$\tilde{P}\approx860$~mW as calculated from Eq.~(\ref{eqn: Ptilde}).   The choice $P_{max} = 2\sqrt{3}\tilde{P}$ made in Fig.~\ref{hysteresis}(a), corresponding to the amount of asymmetry observed in the peak heights when the laser is swept in the two directions, is then equal to 
$P_{max} \approx 3$~W. The maximum circulating power $P_{circ}=P_{trans}/T_2$ that we infer from the transmitted power in Fig.~\ref{hysteresis}(b), on the other hand, is in the vicinity of $P_{circ}\approx 1.5$~W.  The reasons for the discrepancy are not yet understood, but may include a value $T_2$ of the output coupler transmission lower than specified, modification of the transition threshold due to dynamical backaction, or photothermal effects.

Figure~\ref{hysteresis}(c) shows spectra when the input power is doubled to $P_{in}=1$~mW; in this case, the oscillations severely perturb the lineshapes predicted by the static theory alone.
As the input power is raised further, the oscillations become so prominent as to largely obscure the static bistability. Figure~\ref{hysteresis}(d) shows a sweep from red to blue detuning made with an input power of $P_{in}=4$~mW, but at a wavelength of $\lambda= 1539.594$~nm, where the finesse is $F\approx 5600$.  The optically-sprung mechanical oscillation frequency is chirped, and the inset shows that it reaches a value in excess of 1~MHz. 

\section{Conclusion}
In conclusion, we have demonstrated a Fabry-Perot cavity employing a mechanically compliant PhC membrane as one end mirror with a record high finesse of $F=35\,000(500)$.  We measured the finesse in three different ways, and the results of the various measurements are consistent, accounting for a small contribution to the measured cavity linewidth from that of the probe laser.  The finesse is found to have a Lorentzian dependence on wavelength, and the global transmission spectrum reflects a tradeoff between the role of the PhC as gatekeeper for photons seeking to enter the cavity, and its role as a loss mechanism limiting the finesse.   With a cavity linewidth less than half the $\nu_m\approx 426$~kHz frequency of the fundamental mechanical mode, the ``long’’ cavity is firmly in the resolved-sideband regime, making it suitable for ground-state optical cooling.  With a mechanical quality factor in excess of one million, the ``long'' and ``short'' cavities that we construct have single-photon cooperativity values of ${\cal C}_0=6\times10^{-4}$ and ${\cal C}_0=6.6\times10^{-3}$, respectively; for comparison, the systems noted in the classic review paper \cite{aspelmeyer_cavity_2014} with similar mechanical frequencies exhibit cooperativities in the range of ${\cal C}_0=10^{-7}$ to ${\cal C}_0=10^{-4}$. We have found the regime of optomechanical bistability to be easily accessible, as demonstrated by hysteresis in the cavity transmission as the detuning of the injected light is swept.  Experimental results for drive powers even modestly above the bistable threshold are complicated, however, by the oscillation arising from dynamical backaction.  In fact, we have found strong oscillation at an optically-sprung mechanical frequency in excess of 1~MHz, despite the damping provided by the air environment.  We believe these devices have an exciting future in traditional applications of optomechanics, such as cooling, and to this end we are attempting to fabricate membranes incorporating these PhC mirrors within a phononic shield \cite{tsaturyan_ultracoherent_2017} in an effort to improve the mechanical quality factor still further. Finally, we are hopeful that this work will stimulate activity in some of the many fascinating applications of optomechanical bistability that have been proposed since its original demonstration.

\begin{acknowledgments}
F. Z. acknowledges support from the Joint Quantum Institute, NIST/University of Maryland, College Park, MD.
Y. B. acknowledges support from the National Institute of Standards and Technology (NIST), Department of Commerce, USA (70NANB17H247).
This research was performed in part in the NIST Center for Nanoscale Science and Technology Nanofab.

Data underlying the results presented in this paper are not publicly available at this time but may be obtained from the authors upon reasonable request.
\end{acknowledgments}


\bibliography{BistableArxiv}

\end{document}